# GPR signal de-noise method based on variational mode decomposition

**Juncai Xu[1*], Zhenzhong Shen[1], Qingwen Ren[1], Xin Xie[2], and Zhengyu Yang[2]**

[1] *Fundamental Science on Radioactive Geology and Exploration Technology Laboratory, East China Institute of Technology*，*Nanchang, Jiangxi ,330013,China*

[2] *Department of Electrical and Computer Engineering, Northeastern University, Boston, MA USA*

[*]*Corresponding author*

**E-mail:** [1]*xujc@hhu.edu.cn*

**Abstract**:

Compared with traditional empirical mode decomposition (EMD) methods, variational mode decomposition (VMD) has strong theoretical foundation and high operational efficiency. The VMD method is introduced to ground penetrating radar (GPR) signal processing. The characteristics of GPR signals validate the method of signal de-noising based on the VMD principle. The validity and accuracy of the method are further verified via Ricker wavelet and forward model GPR de-noising experiments. The method of VMD is evaluated in comparison with traditional wavelet transform (WT) and EEMD (ensemble EMD) methods. The method is subsequently used to analyze a GPR signal from a practical engineering case. The results show that the method can effectively remove the noise in the GPR data, and can obtain high signal-to-noise ratios (SNR) even under strong background noise.

# 1 Introduction

Ground penetrating radar (GPR) is a rapid and effective means to discover soil composition, widely used in the construction of a project. During the process of GPR signal acquisition, signals are often disturbed by various noises so that acquired GPR signals cannot reflect actual testing results. Therefore, the de-noising process occupies an important role in GPR signal processing and interpretation; it effectively extracts the real signal from the noise background [1, 2].

There are many GPR signal de-noising methods. In the early stages, people often used Fourier transform in the frequency domain for signal filtering. However, the Fourier transform is a method that characterizes a signal's frequency characteristics under the assumption that the signal is stationary, and gives the average frequency characteristic of the analyzed signal over the entire time domain [3]. Therefore, the Fourier transform did not fully characterize non-stationary signals.

Wavelet transform (WT) has also been used to de-noise GPR signals. WT is a kind of time-scale analysis method, which has time-frequency localization characteristics. However, once the base function is determined with WT, it cannot be changed during the process of analysis [4, 5, 6, 7].

To resolve this problem, Huang proposed the empirical mode decomposition (EMD) method which has adaptive characteristics, especially useful for non-linear non-stationary signal processing. EMD discards high-order modal components to achieve a de-noising function [8, 9, 10, 11, 12]. However, the main problem of the EMD method is the lack of a supporting mathematical theory, not to mention others like endpoint effects, modal aliasing, computational problems, and so on.

Variational mode decomposition (VMD) is a new modal decomposition method proposed in recent years, which transfers the acquisition of the signal components to a variational framework. The decomposition of the original signal is realized by constructing and solving constrained variational problems. VMD has a strong supporting mathematical theory, and also a high operational efficiency. It can effectively solve the shortcomings of the conventional EMD method [13-15].

Based on VMD's advantages, the VMD method is introduced to GPR signal processing. According to the characteristics of the GPR signal, we propose the GPR signal analysis method of VMD. This de-noising method of GPR signal is based on the VMD principle. The effectiveness and correctness of the method will be verified by GPR Ricker wavelet and forward modeling experiments. The results will then be compared with those of traditional wavelet transform (WT) and ensemble empirical mode decomposition (EEMD). Finally, the method is applied to GPR signal de-noising in an engineering case example.

# 2 VMD de-noise principle

Actually, the VMD is a variational problem. To minimize the sum of the estimated bandwidths of each mode, we assume that each modal has a finite bandwidth with different center frequencies. Alternating direction multiplier Method is adopted to constantly update the mode and its center frequency, and the mode gradually demodulate to the corresponding baseband. Then the final mode and the corresponding center frequency are extracted.

Assuming that a signal $S_0$ is decomposed into $N$ intrinsic mode function (IMF), the corresponding variational problem's solution can be expressed as follows:

1) The Hilbert transform of each IMF component is used to obtain the analytic signal

$$\left(\delta(t)+\frac{j}{\pi t}\right)*u_k(t) \quad (1)$$

2) The center frequency is estimated for the obtained analytic signal, and the spectrum of each analytical signal transformed to baseband with a frequency shift

$$\left[\left(\delta(t)+\frac{j}{\pi t}\right)*u_k(t)\right]e^{-j\omega_k t} \quad (2)$$

3) The $L^2$ norm of the demodulated signal is calculated and the bandwidth of each mode is estimated. The variational problem is expressed as follows:

$$\begin{cases} \min_{\{u_k\},\{w_k\}}\left\{\sum_k \left\|\partial_t\left[\left(\delta(t)+\frac{j}{\pi t}\right)*u_k(t)\right]e^{-j\omega_k t}\right\|^2\right\} \\ s.t. \sum_k u_k = f \end{cases} \quad (3)$$

For the mentioned variational problem, the quadratic penalty function and Lagrange multiplier are used to transform the problem into an unconstrained problem form

$$L(\{u_k\},\{\omega_k\},\lambda) = \alpha\sum_k\left\|\partial_t\left[\left(\delta(t)+\frac{j}{\pi t}\right)*u_k(t)\right]e^{-j\omega_k t}\right\|_2^2 \\ + \left\|f(t)-\sum_k u_k(t)\right\|_2^2 + \left\langle\lambda(t),f(t)-\sum_k u_k(t)\right\rangle \quad (4)$$

Where $\alpha$ is the penalty factor, and $\lambda(t)$ is the Lagrange multiplier.

Finally, the multiplier alternate direction algorithm is used to solve the unconstrained variational problem of Equation (3).

VMD can be used to decompose the signal into different modes. When the signal contains random noise, a high frequency mode exists in the separated mode. If the high frequency modal component is removed, the random noise will be decreased.

Sample entropy is adopted to determine if higher-order modes are to be preserved. If the modal sample entropy is less than the threshold, the modal will be retained. Otherwise, the modal will be removed. The expression is as follows:

$$\begin{cases} S = \sum_{k=1}^{M} u_k \\ SE_M \leq R \end{cases} \quad (5)$$

Where $SE$ is sample entropy, and $R$ is threshold.

## 3  GPR signal de-noise experiment

In order to verify the effectiveness of the proposed method, we considered two experiments. The methodologies used to analyze were the Rick signal and the forward model.

In general, the containing noise signal is de-noising data minus original noiseless data, and the ratio of signal to noise energy is regarded as signal to noise ratio (SNR). We can use this parameter to measure the de-noising effect.

The formula is as follows:

$$SNR = \frac{\|M\|_2^2}{\|S-M\|_2^2} \quad (6)$$

Where $SNR$ is signal to noise ratio, $\|\ \|_2^2$ is square norm, $M$ is original noiseless data, and $S$ is containing noise data.

### 3.1 Ricker signal analysis

The Ricker signal is the most basic component of a GPR signal. We therefore considered one Ricker signal and taking center frequency to be 50MHz, and time length 1000ns. The initial wave is shown in Fig. 1 (a). Taking the Ricker signal as the object, we take SNR = -13.769, and add random noise into initial wave. The containing noise signal is shown in Fig.1 (b).

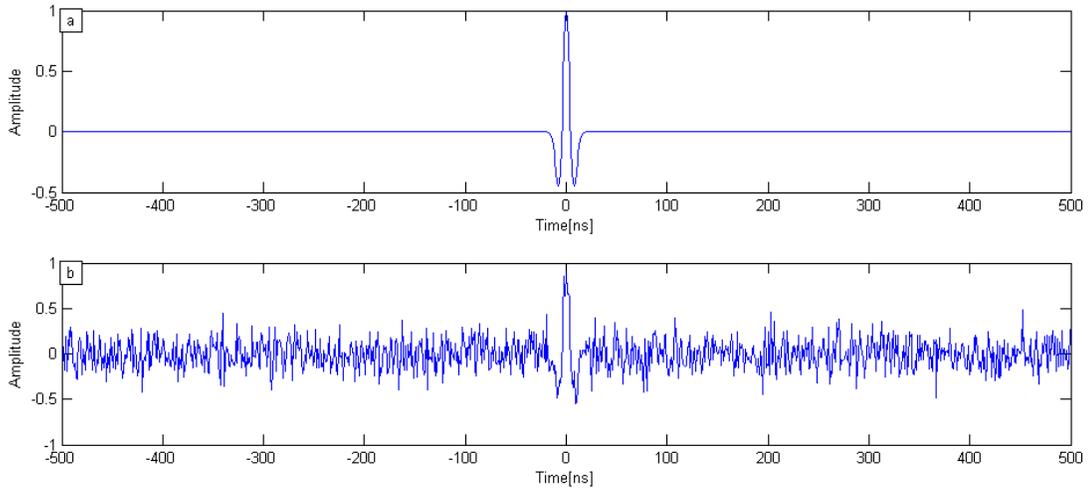

**Fig.1.** Ricker signal (a) Original signal (b) Containing noise signal.

In the de-noising process, we considered three de-noising methods (WT, EEMD and VMD). The three de-noising methods were used to de-noise the signal in Fig.1 (b). The de-noised results are shown in Fig.2.

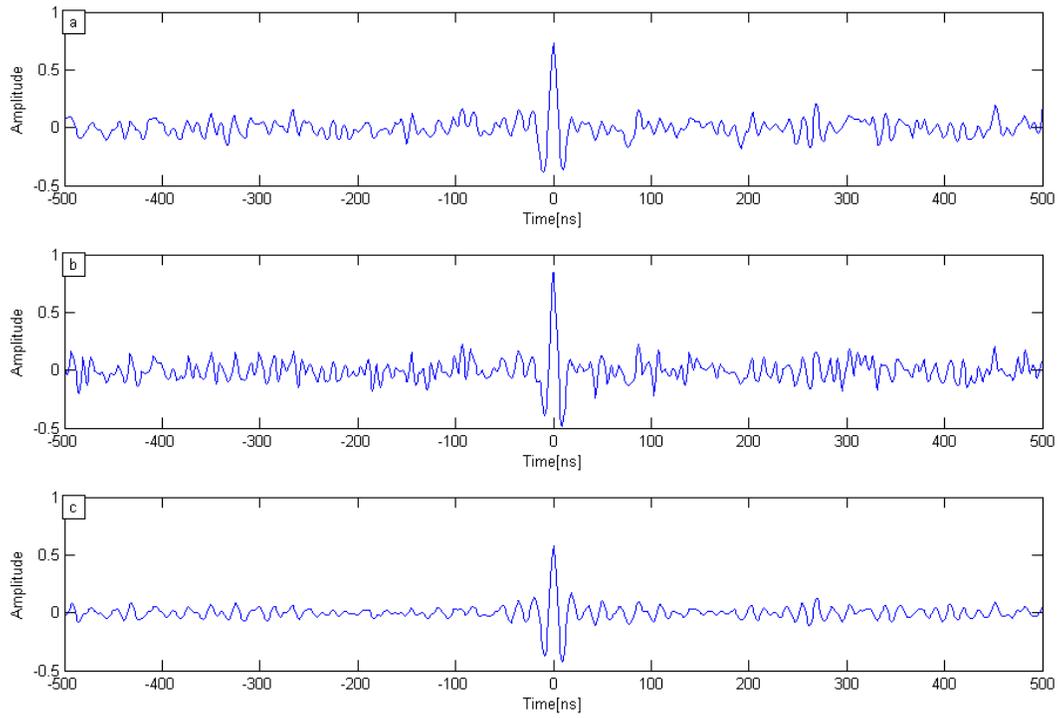

**Fig.2.** Ricker signal de-noise (a) EEMD (b) WT (c) VMD.

According to the de-noising effect in Fig.2, the three methods can effectively suppress the random noise, but the interference component of random noise for VMD as shown in Fig. 2 (c) is obviously lower than the other two. According to Equation (6), the SNR from using each of three different methods are shown in Fig.3.

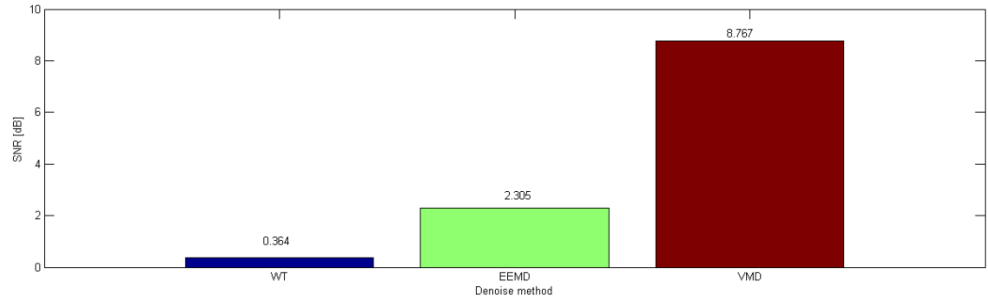

**Fig.3.** Ricker signal SNR with the three de-noise methods.

Based on Fig.3, the de-noise performance of WT is the weakest, the de-noise performance of EEMD method is the second and the de-noise performance of VMD is the strongest among the three methods. Therefore, the VMD de-noising method is superior to WT and EEMD in GPR wavelet signal de-noising.

### 3.2   Analysis of forward synthetic data

Here we used a two-layer structure model. The uppermost layer being air and the bottom layer dry sand. A circular void is placed in the dry sand layer. The model size and location are shown in Fig. 4. Both the concrete and sand layers have thicknesses of 0.15m, while the internal holes diameter size is 0.075m. The GPR signal of the model can be obtained through numerical simulation.

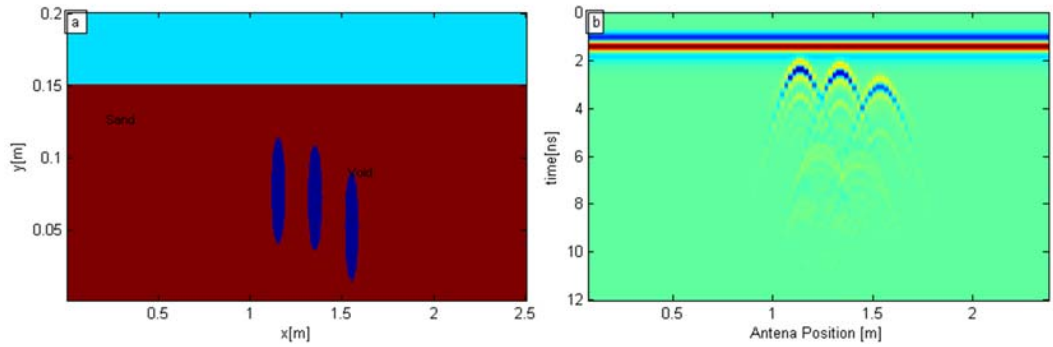

**Fig.4.** GPR forward model (a) Schematic of geometric model (b) Forward simulation signal of GPR.

In the forward process, the parameters are dry sand permittivity (3), dry sand conductivity (0.0001 S/m), air permittivity (1), and air conductivity (1.0e-10 S/m). The set calculation area measures 2.5 m × 0.15 m, the computational grid size is 0.0025 m, and the time window is 12 ns. Ricker wave is employed as excitation, the dominant frequency is at 900 MHz, the transceiver antenna distance is 0.025 m, and the analogue sampling is 125 traces. The FDTD method is performed for GPR signals of the time series for all traces. The measurement line direction is the x-axis, and time is the y-axis in ordinate. The time-series signal of 125 traces is synthesized, and the result is displayed using image mode in Fig.4 (b).

We take SNR=-11.823dB and add random noise into the simulation signal. Fig.5 (b) is the GPR signal after adding noise. The radar image in Fig.5 (b) shows that the event and reflection arcs are relatively vague. It is difficult to clearly distinguish the internal structures.

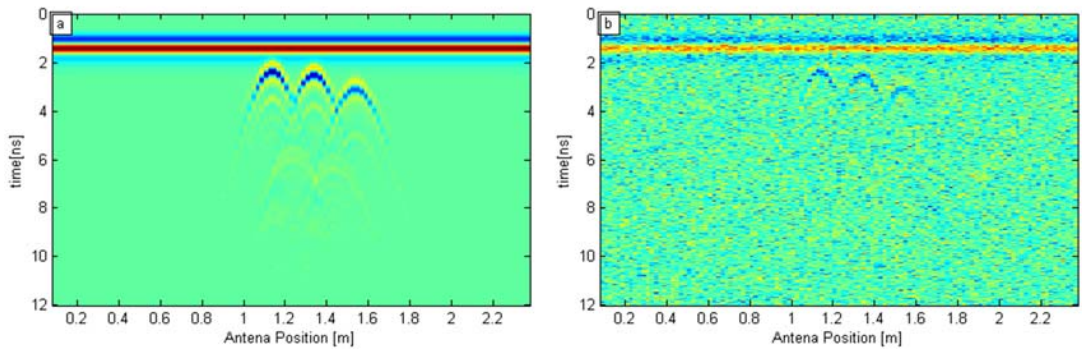

**Fig.5.** GPR forward model experiment (a) Original signal. (b) Addition of white noise signal.

We take containing the noise signal in Fig.5 (b) as the studying object, and consider WT, EEMD and VMD to de-noise. The three methods are used to deal with the containing noise signal. The processed results are shown in Fig.6.

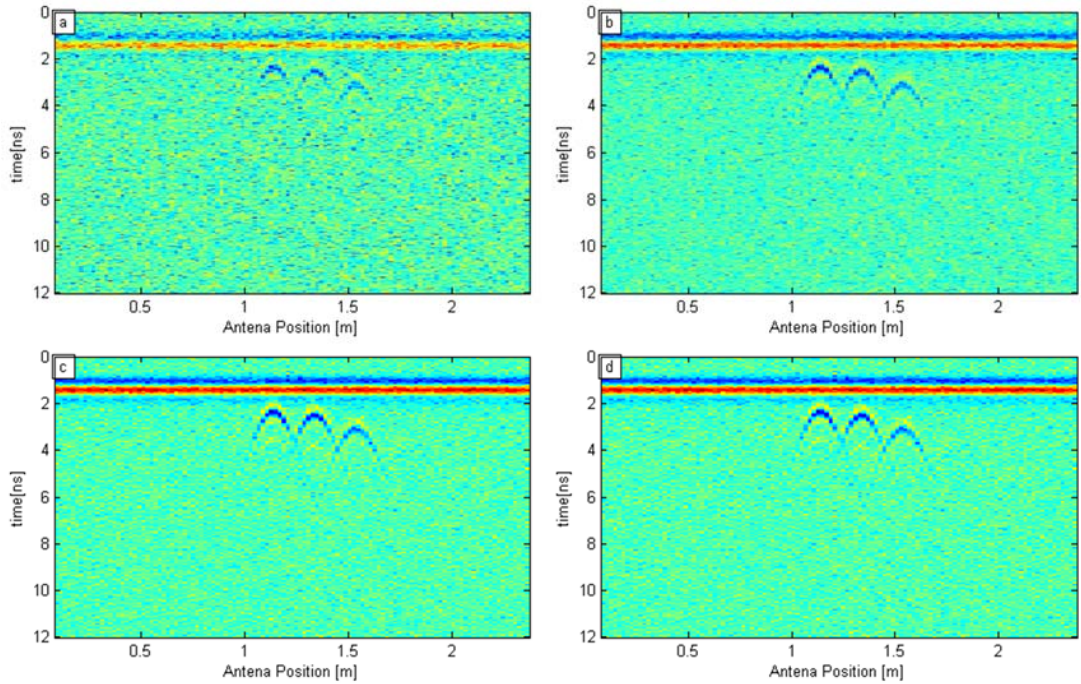

**Fig.6.** The de-noise result with the three methods (a) Containing noise signal (b) WT (c) EEMD (d) VMD.

Based on the comparison of Fig.6, the three methods can effectively remove the random noise from the GPR signal. In comparison with the de-noise effect of the other two methods, the signal with VMD de-noising has the most clarity, however the signal with WT de-noising is the least clear.

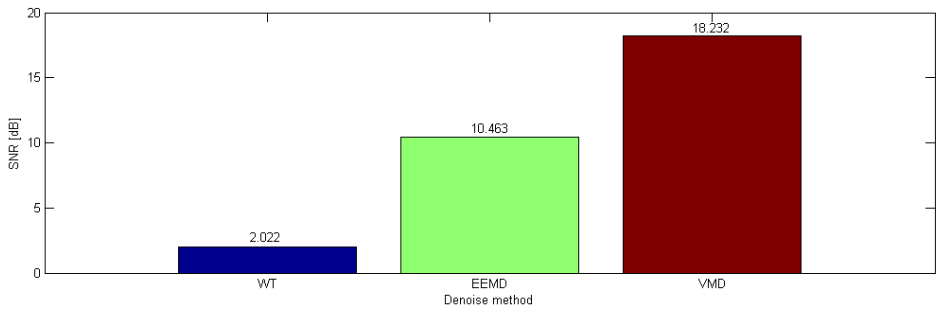

**Fig.7.** Forward GPR signal SNR with the three de-noising methods.

The de-noise performance of the three methods is shown in Fig.7. Based on Fig. 7, the de-noise performance of the WT is the lowest, the de-noise performance of the EEMD is the second and the de-noise performance of VMD is the best. The result is consistent with Section 3.1.

## 4　Engineering application

The GPR method was applied to investigate the location of a water plant in Longnan County, Jiangxi Province. The survey line was located on the river site. The detection device used was a LTD2100 G PR. We chose a 100 MHz antenna to explore the site, as shown in Fig.8 (a). The detection section length was 128 m, and the track step was 0.50 m. The cross section had 256 trace records, and each trace contained 400 sampling points. Fig.8 (b) shows the survey records of the GPR site.

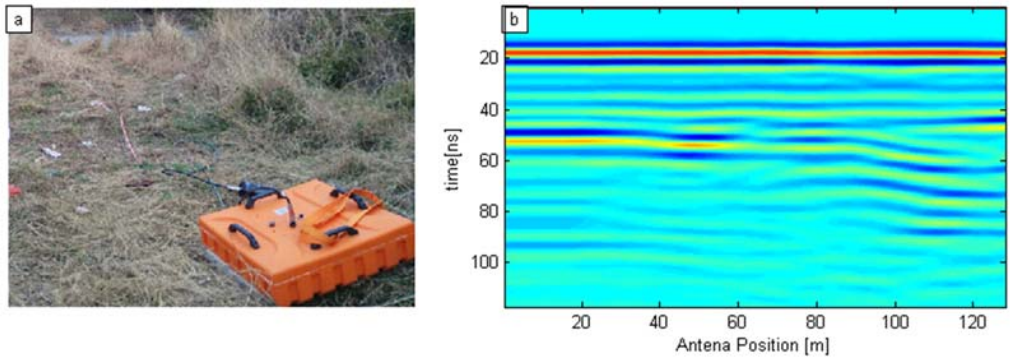

**Fig.8.** GPR geological exploration: (a) On the spot GPR measurement and (b) GPR section records.

In the process of removing noise via VMD, the SNR of signal was set to -4.372 dB. We added the original signal to the strong interference and the signal containing strong interference is shown in Fig.9 (b). In the strong interference background, the reflected strength from layers is very vague. It is difficult to distinguish the formation depth.

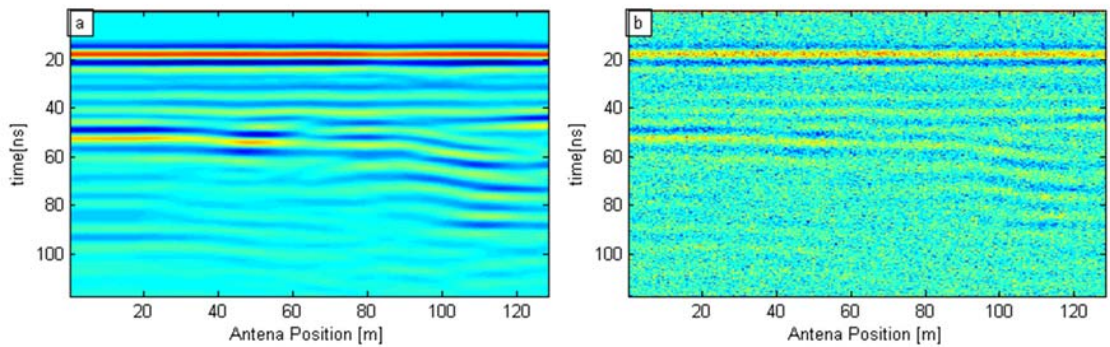

**Fig.9.** GPR signals (a) Original signal. (b) Containing white noise signal.

In order to remove the influence of noise interference, the VMD method is used to de-noise the GPR signal in Fig. 9 (b).The GPR signal after de-noising is obtained with VMD. Fig.10 shows the image of GPR before and after de-noising.

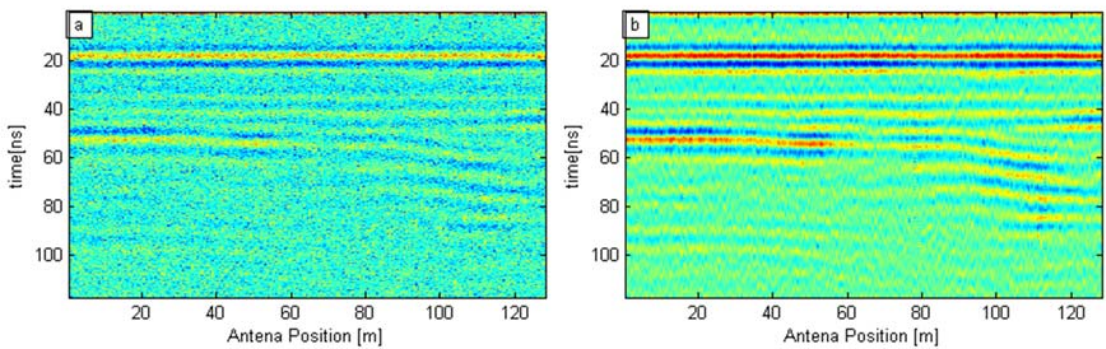

**Fig.10.** GPR signal with VMD de-noise (a) Containing white noise signal. (b) Signal after de-noised.

Based on the analysis results in Fig.10, the radar images after de-noise becomes clear and the stratigraphic boundaries can be identified. SNR increased to 14.677dB; it was greatly improved by VMD. Therefore, the VMD method can significantly filter out interference effects.

## 5 Conclusions

Based on the principle of VMD, the VMD method is introduced to GPR signal analysis. Focusing on the characteristics of GPR signal, a de-noising method is proposed that utilizes VMD. In the de-noise experiments, the VMD method was used to de-noise GPR Ricker wavelet and forward model signals, and its de-noising effect is compared to that of the WT and EEMD methods. VMD is also applied to GPR signal in an exploration engineering practical case. Through the research and analysis, the following conclusions are drawn:

(1) The VMD method can remove the noise in the radar signal and have high de-noise performances;

(2) In strong noise backgrounds, the VMD method can effectively remove the noise from the GPR signal and obtain a high SNR;

(3) VMD method can obtain better SNRs than other de-noising methods, and this has significant advantages.

**Acknowledgment**

This research was funded by the Open Research Fund of the Fundamental Science on Radioactive Geology and Exploration Technology Laboratory (Grant No. RGET1502) and the National Natural Science Foundation of China (Grant No. 51279054). The author is grateful to Wang Hailong from the Chongqing Test Center of Geology and Minerals for providing the GPR practical detection data used in this study.